%% file: QS-v2.tex
\documentclass[letterpaper, 11pt]{article}
\title{\Large Quantum Simulation of Simple Many-Body Dynamics}
\author{\normalsize Yale Fan}
\date{\normalsize \today}
\input{Qcircuit}
\input{xy}
\input{paperformats}
\xyoption{all}
\usepackage{amsmath}
\usepackage{graphicx}
\arxivformat

\begin{document}
\maketitle
\begin{abstract}
Quantum computers could potentially simulate the dynamics of systems such as polyatomic molecules on a much larger scale than classical computers.  We investigate a general quantum computational algorithm that simulates the time evolution of an arbitrary non-relativistic, Coulombic many-body system in three dimensions, considering only spatial degrees of freedom.  We use a simple discretized model of Schr\"odinger evolution and discuss detailed constructions of the operators necessary to realize the scheme of Wiesner and Zalka.  The algorithm is simulated numerically for small test cases, and its outputs are found to be in good agreement with analytical solutions. 
\end{abstract}

\section{Introduction}

Quantum computers could provide an exponential speedup over classical methods of simulating quantum mechanical systems \cite{Feynman}.  Wiesner \cite{Wiesner} and Zalka \cite{Zalka} have sketched a basic theory of such simulations that includes discretization techniques and decomposition of time evolution operators.  Other approaches in the literature include the quantum lattice gas automaton (QLGA) \cite{Generic, Lattice, Yepez} and various methods of decomposing fermionic Hamiltonians into spin operators via the Jordan-Wigner transformation \cite{Abrams, Ortiz, Ovrum, Lanyon}.

However, work based on the first approach has generally focused on bounds for relevant time or space complexities rather than the explicit form of the requisite operators \cite{Giuliano, Kassal, NC}, as well as on the problem of calculating energy spectra \cite{Aspuru-Guzik, Lanyon}.  Conversely, in this note, we expand on Wiesner and Zalka's approach to compute position-dependent wavefunctions.  We propose detailed constructions of the operators (quantum gates) for such simulations and numerical tests of the algorithm described.

Given the initial state of some number of particles confined to a finite volume in three-dimensional space, we aim to compute the wavefunction of the system after an arbitrary period of time.  Our simplified model uses only the non-relativistic Coulomb Hamiltonian, whose terms include kinetic energies and electric potentials, and neglects spin-orbit effects.

\section{Method} \label{method}

\subsection{Hamiltonian}

The Hamiltonian $\hat{H}$ is a Hermitian operator acting on states in a Hilbert space.  For quantum computation, the Hilbert space in question is finite-dimensional.  We will consider primarily the Coulomb Hamiltonian $\hat{H} = \hat{T} + \hat{U}$, which consists of terms of the form
\[
\hat{T} = -\sum_i \frac{\hbar^2}{2m_i}\nabla_i^2 \mbox{ and } \hat{U} = \frac{1}{4\pi\epsilon_0} \sum_i\sum_{j > i} \frac{q_i q_j}{\| \mathbf{r}_i - \mathbf{r}_j \|}.
\]
For a molecular system,
\begin{equation}
\label{molecular Hamiltonian}
\hat{H} = \hat{T}_e + \hat{T}_n + \hat{U}_{ee} + \hat{U}_{nn} + \hat{U}_{en}
\end{equation}
where the $\hat{U}$ terms represent various interactions between electrons and nuclei.  One of our goals is to make such continuous operators precise in the context of quantum computing: what does the single-particle momentum operator $-i\hbar\nabla$ mean as a finite-dimensional unitary matrix and a quantum gate?

We use the following simulation parameters to determine the dimension of the Hilbert space and the evolution of $\psi$: $T$, the total time over which to simulate; $N_t$, the number of small time increments; $L$, the bound on the dimensions of the system; and $n$, a parameter chosen to reflect our desired precision.

\subsection{Discretization} \label{discretization}

Wiesner \cite{Wiesner} and Zalka \cite{Zalka} have outlined algorithms for simulating quantum many-body systems with the common element of diagonalization using the quantum Fourier transform (QFT).  Expanding on their framework, we explicitly construct the operators needed to carry out these simulations.

We first review the common encoding scheme for Cartesian coordinates in three spatial dimensions.  Let all particles be confined to a finite volume ($0 \leq x, y, z \leq L$).  We divide the intervals along each coordinate axis into $2^n$ subintervals of length $\delta = L/2^n$.  The volume is hence partitioned into many subvolumes (position basis states) where each amplitude of the state vector of a particle in the system specifies the probability that it will be found in a subvolume.  The system wavefunction is a tensor product of discretized single-particle wavefunctions, or a sum thereof, encoded in a register of qubits.  We approximate a continuous single-particle wavefunction $\psi(\mathbf{r}, t)$ by the state vector
\begin{equation}
\label{state vector}
|\psi(\mathbf{r}, t)\rangle = \sum_{i=0}^{2^n-1}\sum_{j=0}^{2^n-1}\sum_{k=0}^{2^n-1} \psi(\mathbf{r}_{ijk}, t)|ijk\rangle,
\end{equation}
normalized by a factor of $(\sum_{i,j,k} |\psi(\mathbf{r}_{ijk}, t)|^2)^{-1/2}$, where
\begin{equation}
\label{r}
\mathbf{r}_{ijk} = (r_x, r_y, r_z) = \delta\left(i+1/2, j+1/2, k+1/2\right).
\end{equation}
The one-dimensional position basis states $|i\rangle$, $|j\rangle$, and $|k\rangle$ range over discrete positions on the $x$, $y$, and $z$ axes, respectively, while $|ijk\rangle$ denotes the tensored state $|i\rangle|j\rangle|k\rangle$.  We assign the amplitude associated with each subvolume to be the amplitude of $\psi$ at its center.  Each particle requires a register of $n$ qubits for each of the $x$, $y$, and $z$ intervals.  The algorithm thus requires $3nN$ total qubits, $N$ being the number of particles.

The particles are assumed spinless, but that leaves the question of distinguishability.  Our algorithm is fully compatible with simulating indistinguishable fermions if one adds ancillary qubit registers to allow for the initialization of an antisymmetric spatial wavefunction via the algorithm of Abrams and Lloyd \cite{Abrams}.  In our notation, this requires $O((3nN)^2)$ operations given $N$ single-particle states.  This procedure also suffices (and requires fewer operations) for symmetrization --- for example, when dealing with bosonic nuclei.  Our main focus here is the form of the operators, so we will not pursue this point further. 

From the Schr\"odinger equation $\hat{H} \psi = i\hbar\partial_t \psi$, we have the time evolution
\begin{equation}
\label{time evolution}
|\psi(\mathbf{r}, t_0 + \epsilon)\rangle = e^{-\frac{i}{\hbar}\hat{H}\epsilon}|\psi(\mathbf{r}, t_0)\rangle.
\end{equation}
For very small time scales $\epsilon$, we can apply the Trotter decomposition \cite{NC} to construct gates for each term of the Hamiltonian separately and apply them sequentially as unitary time evolution operators:
\begin{equation}
\label{time evolution operator}
e^{-\frac{i}{\hbar}\hat{H}\epsilon} = e^{-\frac{i}{\hbar}\hat{U}\epsilon} e^{-\frac{i}{\hbar}\hat{T}\epsilon} + O(\epsilon^2).
\end{equation}
To carry out the algorithm, we choose a time interval $T$ over which to simulate and partition it into $N_t$ small intervals $\epsilon = T/N_t$.  At each of the $N_t$ time steps, we apply the operator \eqref{time evolution operator} to the state vector of our system in the position basis.

Na\"ively, one can repeat the entire algorithm many times and make subsequent measurements of every qubit to iteratively determine a probability distribution for the particle configurations, thus allowing one to predict the overall spatial density at the end of the desired time interval.  The procedure is identical to that required for the algorithm of Abrams and Lloyd \cite{Abrams} for simulating fermions in the first-quantized representation, namely constructing a histogram of configurations obtained from the repeated measurements.


\subsection{Construction of Kinetic Energy Operators} \label{kinetic}

It is commonly noted \cite{Wiesner, Zalka, Giuliano, Kassal, NC} that implementing the kinetic energy operator by Fourier transforming to the momentum basis and back reduces the time evolution operators to diagonal operators and the QFT.  However, the form of the kinetic energy operator in the model of Wiesner and Zalka has not, to the author's knowledge, been previously addressed.

To this end, we determine the form of the momentum operator matrix $\hat{P}$ by analogy with the continuous case.  First, consider the $-i\hbar\nabla$ operator for a single particle in one dimension: i.e., along an $x$-interval of length $L$ divided into $2^n$ subintervals of length $\delta$.  To obtain approximations for the derivative of the discrete-valued wavefunction $\psi$ at each point $k$ ($0\leq k < 2^n$) corresponding to a position basis state, define
\[
D_k^+ = \frac{\psi(k + 1) - \psi(k)}{\delta}, \hspace{2 mm} D_k^- = \frac{\psi(k) - \psi(k - 1)}{\delta}, \hspace{2 mm} D_k^\textrm{ave} = \frac{\psi(k + 1) - \psi(k - 1)}{2\delta}
\]
so that to the best approximation,
\begin{eqnarray*}
D_x\psi(0) &=& D_0^+ + O(\delta), \\
D_x\psi(2^n - 1) &=& D_{2^n - 1}^- + O(\delta), \mbox{ and} \\
D_x\psi(k) &=& D_k^\textrm{ave} + O(\delta^2) \hspace{2 mm} (0 < k < 2^n - 1).
\end{eqnarray*}
The action of the ``discrete'' derivative operator is thus described by
\[
\sum_{k=0}^{2^n-1}\psi(k)|k\rangle \stackrel{\hat{D}_x}{\longmapsto} D_0^+ |0\rangle + \sum_{k=1}^{2^n-2}D_k^\textrm{ave}|k\rangle + D_{2^n - 1}^- |2^n - 1\rangle,
\]
or in matrix form as
\begin{equation}
\label{derivative matrix}
\hat{D}_x = \frac{1}{2\delta} 
\left(\begin{array}{cccccc}
-2 & 2 & 0 & \cdots & 0 & 0 \\
-1 & 0 & 1 & \cdots & 0 & 0 \\
0 & -1 & 0 & \ddots & \vdots & \vdots \\
\vdots & \vdots & \ddots & \ddots & 1 & 0 \\
0 & 0 & \cdots & -1 & 0 & 1 \\
0 & 0 & \cdots & 0 & -2 & 2
\end{array}\right).
\end{equation}
In addition, we require that the matrix $-\frac{i}{\hbar}(\hat{P}^2/2M)\epsilon$ be skew-Hermitian so that its exponential is unitary.  Hence $\hat{P}^2$ must be symmetric, and we can make it so by sacrificing the approximations for $\hat{D}_x$ at the endpoints.  The corresponding momentum operator is
\begin{equation}
\label{P}
\hat{P} = -i\hbar\hat{D}_x \approx -\frac{i\hbar}{2\delta}
\left(\begin{array}{cccc}
0 & 1 & \cdots & 0 \\
-1 & 0 & \ddots & \vdots \\
\vdots & \ddots & \ddots & 1 \\
0 & \cdots & -1 & 0
\end{array}\right).
\end{equation}
The form of the kinetic energy operator follows immediately, and extending the kinetic energy operator to three dimensions is as simple as applying it to the qubits encoding the $x$-, $y$-, and $z$-coordinates of each particle separately.  Since this construction takes place in the position basis, its implementation does not require a QFT.  We next discuss two methods of implementing the kinetic energy term of the time evolution operator $e^{-\frac{i}{\hbar}\hat{T}\epsilon}$ as a set of quantum gates.

As a first method, note that
\[
\hat{P}^2 = -\frac{\hbar^2}{4\delta^2}\left(-|0\rangle\langle 0| + \sum_{i=1}^{2^n-2} \hat{P}_i - |2^n-1\rangle\langle 2^n-1|\right)
\]
where $\hat{P}_i = |i-1\rangle\langle i+1| - 2|i\rangle\langle i| + |i+1\rangle\langle i-1|$.  Let $\xi = \frac{i\hbar\epsilon}{8M\delta^2}$.  The Trotter formula gives
\begin{equation}
\label{KE Trotter}
e^{-\frac{i}{\hbar}\left(\frac{\hat{P}^2}{2M}\right)\epsilon} = e^{-\xi |0\rangle\langle 0|} \left(\prod_i e^{\xi\hat{P}_i}\right) e^{-\xi |2^n-1\rangle\langle 2^n-1|} + O(\xi^2)
\end{equation}
where $e^{-\xi |0\rangle\langle 0|}$ and $e^{-\xi |2^n-1\rangle\langle 2^n-1|}$ are both single-qubit phase rotation operators.  Furthermore, since each of the $\hat{P}_i$ is a very sparse matrix, the $e^{\xi\hat{P}_i}$ are block diagonal matrices representing two-qubit operations: $e^{\xi\hat{P}_i} = \operatorname{diag}(I_{i-1}, M_P, I_{2^n-2-i})$ where
\[
M_P = \exp
\left(\begin{array}{ccc}
0 & 0 & \xi \\
0 & -2\xi & 0 \\
\xi & 0 & 0
\end{array}\right)
\]
and $I_m$ denotes the $m\times m$ identity matrix.  Hence we have decomposed the kinetic energy operator entirely into controlled one- and two-qubit gates.  Since $2^n$ such gates act on each $n$-qubit register representing position basis states along one coordinate axis, $3N2^n$ one- and two-qubit gates are needed to simulate the kinetic time evolution operator of the entire system.  This gate complexity is linear in the number of subintervals along each axis, or the ``absolute'' precision $\widetilde{N} = 2^n$.

A second, and more robust, method is to realize that in the continuum limit, $\hat{P}$ will be exactly diagonalized by the QFT, as follows: letting $D$ be the dimension of the Hilbert space, $F$ be the QFT of dimension $D$, and $\alpha = -i\hbar/2\delta$, we can write
\[
\hat{P} = \alpha \sum_{x=0}^{D-2} (|x\rangle\langle x+1| - |x+1\rangle\langle x|).
\]
Applying $F = \frac{1}{\sqrt{N}}\sum_{j,k=0}^{N-1} e^{i2\pi jk/N} |j\rangle\langle k|$ and using $\langle j|k\rangle = \delta_{jk}$ for simplification shows that
\begin{equation}
F\hat{P}F^\dag = \frac{\alpha}{D} \sum_{j=0}^{D-1} \sum_{k=0}^{D-1} (e^{-i2\pi k/D} S_1(j, k) - e^{i2\pi k/D} S_2(j, k)) |j\rangle\langle k|
\end{equation}
where
\[
S_1(j, k) = \sum_{p=0}^{D-2} e^{i2\pi p(j-k)/D} \mbox{ and } S_2(j, k) = \sum_{p=1}^{D-1} e^{i2\pi p(j-k)/D}.
\]
Summing the geometric series gives $S_1(j, k) = -e^{i2\pi(j-k)(D-1)/D}$ and $S_2(j, k) = -1$ if $j\neq k$.  Consequently, as $D\rightarrow\infty$, $S_1(j, k) \rightarrow -1$ and the non-diagonal entries approach
\begin{equation}
\frac{\alpha}{D}(e^{i2\pi k/D} - e^{-i2\pi k/D}) = \frac{\alpha}{D}\left(2i\sin\frac{2\pi k}{D}\right) \rightarrow 0.
\end{equation}
If $j = k$, $S_1(j, k) = S_2(j, k) = D - 1$, so that as $D\rightarrow\infty$, the $k^\mathrm{th}$ diagonal entry becomes
\begin{equation}
\label{diagonal entries}
\frac{D-1}{D}\alpha(e^{-i2\pi k/D} - e^{i2\pi k/D}) \rightarrow -2i\alpha\sin\frac{2\pi k}{D} = -\frac{\hbar}{\delta}\sin\frac{2\pi k}{D}.
\end{equation}
Thus $F\hat{P}F^\dag$ is a diagonal matrix with real eigenvalues in the continuum limit, as expected physically.  Now that we have explicitly determined these eigenvalues, we can construct the kinetic energy operator from QFTs and a diagonal $\hat{P}$ matrix with entries of the form \eqref{diagonal entries} in the momentum basis.  Once again, an unoptimized method of implementing the associated time evolution operators requires only $3N(2^n + O(n^2))$ gates --- taking into account both the QFT and the diagonal component --- which is linear in both the number of particles and the number of subintervals along each coordinate axis.

\subsection{Construction of Potential Energy Operators}

The construction of the potential energy operators is generally straightforward in the position basis.  For example, the Coulomb potential operator for a molecular system has the following physical parameters: $N_e, \mathbf{r}_i$ (resp.\ $N_n, \mathbf{R}_i$) for the number and positions of the electrons (resp.\ nuclei) and $Z_i, M_i$ for the atomic numbers and masses of the nuclei.  Take $\mathbf{r}_p \equiv \mathbf{r}_{i_p j_p k_p}$ (resp.\ $\mathbf{R}_p \equiv \mathbf{R}_{I_p J_p K_p}$) as shorthand, where $i_p j_p k_p$ (resp.\ $I_p J_p K_p$) is a bit string corresponding to the binary representation of some index describing a position basis state, or subvolume, occupied by the $p^\mathrm{th}$ electron (resp.\ nucleus).  Recall that the position vector $\mathbf{r}_{ijk}$ is defined in equation \eqref{r}.  Also, let $A[m]$ denote the $m^\mathrm{th}$ diagonal entry of matrix $A$.

$\hat{U}_{ee}$ can be seen as an operator on the state of the entire system of electrons, or as a $2^{3nN_e} \times 2^{3nN_e}$ diagonal matrix whose entries give the total potential energy due to any of the $2^{3nN_e}$ allowed configurations of electrons.  If
\[
|m\rangle = \bigotimes_{s=1}^{N_e}|i_s j_s k_s\rangle
\]
represents one of the many possible electron configurations, then
\begin{equation}
\label{U_ee entries}
\hat{U}_{ee}[m] = \frac{e^2}{4\pi\epsilon_0} \sum_{p=1}^{N_e} \sum_{q=p+1}^{N_e} \frac{1}{\| \mathbf{r}_p - \mathbf{r}_q \|}.
\end{equation}
Analogously, $\hat{U}_{nn}$ acts on the state vector of the entire system of nuclei and $\hat{U}_{en}$ on all particles.  For those diagonal entries that represent configurations of the system in which multiple particles are found in the same ``box'' of volume (i.e., when $|i_p j_p k_p\rangle = |i_q j_q k_q\rangle$ and hence $\mathbf{r}_p = \mathbf{r}_q$ for some $p, q$), we can set the potential to some approximate large, but finite, value.  For example, in the case of two electrons, we can replace the undefined term $e^2/4\pi\epsilon_0 \| \mathbf{r}_p - \mathbf{r}_q \|$ with the approximation $e^2/4\pi\epsilon_0 \delta$.

\subsection{Complexity of Potential Energy Operators}

It is unknown whether nonlocal potential energy operators such as these can be implemented in a polynomial number of resources; Lloyd demonstrated efficiently-scaling implementations for local interactions \cite{Lloyd}.  Benenti and Strini \cite{Giuliano} have shown that the time evolution operators corresponding to a harmonic oscillator potential do scale polynomially with the number of qubits, and their argument is easily extended to power-law potentials of the form $V(x) = ax^r$ where $r$ is a positive integer.  The difficulty with the Coulomb potential lies in $r$ being negative, which prevents easy factorization of the operators into products of phase-shift gates acting on a fixed number of qubits, even after introducing ancillary qubits.  However, it may be possible to implement these nonlocal potential operators in polynomial time if we choose to sacrifice some additional precision in computing the values of our potential energies.  Polynomial scaling is important because it is necessary to first calculate the potential energies \eqref{U_ee entries} on a classical computer before implementing the corresponding phase rotations in quantum gates.

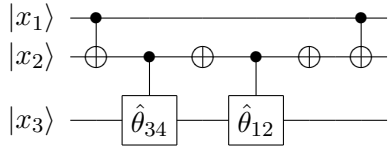
\begin{figure}[htb]
\centerline{\Qcircuit @C=0.5em @R=1em {
\lstick{|x_1\rangle} & \ctrl{1} & \qw & \qw & \qw & \qw & \qw & \ctrl{1} & \qw \\
\lstick{|x_2\rangle} & \targ & \ctrl{1} & \targ & \ctrl{1} & \targ & \qw & \targ & \qw \\
\lstick{|x_3\rangle} & \qw & \gate{\hat{\theta}_{34}} & \qw & \gate{\hat{\theta}_{12}} & \qw & \qw & \qw & \qw }}
\caption{Efficient implementation of a diagonal operator. \label{efficient}}
\end{figure}

First, we can utilize redundancy to implement the potential energy operators more efficiently.  As an example, consider the three-qubit phase shift operator
\[
\operatorname{diag}(e^{i\theta_1}, e^{i\theta_2}, e^{i\theta_3}, e^{i\theta_4}, e^{i\theta_3}, e^{i\theta_4}, e^{i\theta_1}, e^{i\theta_2})
\]
with ``redundant'' entries.  The na\"ive method of constructing this diagonal operator is to use the first two qubits as controls for phase shift operators on the third qubit, resulting in four gates acting on the third qubit.  A more efficient scheme is shown in Figure \ref{efficient}, using the single-qubit gates
\[
\hat{\theta}_{12} =
\left(\begin{array}{cc}
e^{i\theta_1} & 0 \\
0 & e^{i\theta_2}
\end{array}\right)
\mbox{ and } \hat{\theta}_{34} =
\left(\begin{array}{cc}
e^{i\theta_3} & 0 \\
0 & e^{i\theta_4}
\end{array}\right).
\]
Because there are many configurations of a molecular system with the same potential energy (up to a precision), the potential energy operators also have many redundant entries.  In fact, let $\hat{U}$ be one of $\hat{U}_{ee}$, $\hat{U}_{nn}$, and $\hat{U}_{en}$, and let the corresponding $N$ be $N_e$, $N_n$, or $N_e + N_n$.  Then $\hat{U}[x] = \hat{U}[2^{3nN} - 1 - x]$ for $0\leq x\leq 2^{3nN-1} - 1$, meaning that the potential energy operators are symmetric about the antidiagonal.  This is easily seen by noting that the indices of the states $|x\rangle$ and $|2^{3nN} - 1 - x\rangle$ are bitwise complements, meaning that by symmetry, they represent reflected versions of the same molecular configuration.  This observation reduces the number of gates needed by a factor of two, and similar optimizations could lead to much more efficient implementations of the potential energy operators.

Second, because we would like to simulate arbitrarily large systems, the following variables are expensive: $N_e$, $N_n$ (in addition to the atomic numbers $Z_i$), and the absolute precision $\widetilde{N}$ where $n = \log_2 \widetilde{N}$.  We have already seen that the kinetic energy operators scale linearly with $\widetilde{N}$.  The number of distinct values of the potential energy of the system can scale polynomially if we restrict our calculations to some appropriate precision $\Delta U$.  Then, despite the exponential number of particle configurations, their potential energies only assume a polynomial number of values, perhaps allowing us to employ redundancy to implement the potential energy operators efficiently.

To see how this might work, let $e' = e^2/4\pi\epsilon_0$.  The extreme values of the potential energy of the entire system might go like
\[
U_\mathrm{max} = \frac{e'}{\delta} \left(\frac{N_e(N_e - 1)}{2} + \sum_i \sum_{j > i} Z_i Z_j\right) \mbox{ and } U_\mathrm{min} = -\frac{e'}{\delta} \sum_i Z_i.
\]
We would like to determine some increment $\Delta U$ so that the number of possible potential energies of the system between $U_\mathrm{min}$ and $U_\mathrm{max}$, to a precision of $\Delta U$, is $(U_\mathrm{max} - U_\mathrm{min})/\Delta U$ and becomes polynomial in the ``expensive'' variables.  The smallest possible change in potential energy between two different configurations of the particles occurs roughly when two electrons are separated by a large distance and one of them is very slightly displaced.  The maximum distance between the two electrons is of order $L$, and the minimum displacement is of order $\delta$.  Let the initial $\mathbf{r}$ vector between the electrons have length $L$ and let the final one have length $L'$.  The minimum change in distance $L' - L$ occurs when the displacement vector of length $\delta$ makes approximately a right angle with the initial $\mathbf{r}$ vector, so that $L' = \sqrt{L^2 + \delta^2} \approx L + \delta^2/2L$.  This assumption is justified because the angle $\theta$ between the initial $\mathbf{r}$ vector and the displacement vector for which $L' = L$ is such that $\cos\theta = \delta/2L \ll 1$, so the angle that minimizes $L' - L$ can be made arbitrarily close to $\pi/2$.  Then the minimum change in potential energy due to this small displacement is
\[
\Delta U = e'\left(\frac{1}{L} - \frac{1}{L'}\right) \approx e'\frac{\delta^2}{2L^3},
\]
whence 
\begin{equation}
\frac{U_\mathrm{max} - U_\mathrm{min}}{\Delta U} \approx \frac{2L^3}{\delta^3}P(N_e, Z_1, \ldots, Z_{N_n}) = \widetilde{N}^3 \widetilde{P}(N_e, Z_1, \ldots, Z_{N_n})
\end{equation}
where $\widetilde{P}(N_e, Z_1, \ldots, Z_{N_n})$ is some polynomial expression of the arguments with constant factors absorbed into it.  Hence it may be possible to construct the potential energy operators in a number of gates polynomial in $N_e$, the $Z_i$, and $\widetilde{N}$ as long as we restrict the precision of the potential energy values to $\Delta U$ and utilize redundancy effectively.

\section{Numerical Tests}

We illustrate the use of these operators in an actual algorithm with two examples.

\subsection{Order of Error: Particle in a Box}

We first numerically demonstrate the accuracy of our approximation for $e^{-\frac{i}{\hbar}\hat{T}\epsilon}$ by using it to determine a time evolution for a particle in a one-dimensional infinite potential well of length $L$, hence dropping the Coulombic term from the Hamiltonian.  The goal here is to determine the order of error of our approximations by simulating the simplest possible system.  We enforce the boundary conditions by introducing a potential energy operator that takes an arbitrarily large value at the endpoints.

\begin{figure}[htb]
\begin{center}
\vspace{5 mm}
$T = 1\times 10^{-3}$ \hspace{2.5 cm} $T = 2\times 10^{-3}$ \hspace{2.5 cm} $T = 3\times 10^{-3}$ \\
\includegraphics[scale=0.39]{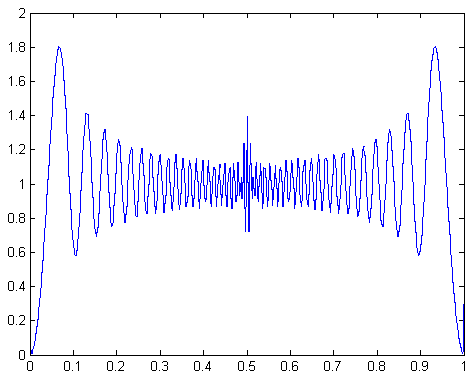} \includegraphics[scale=0.39]{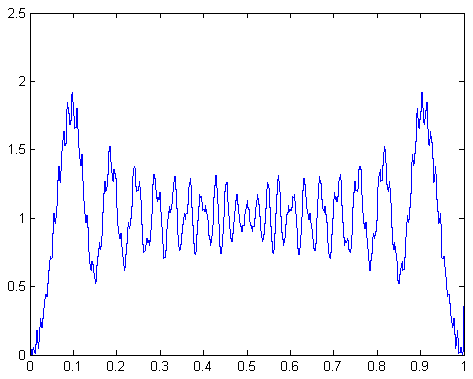} \includegraphics[scale=0.39]{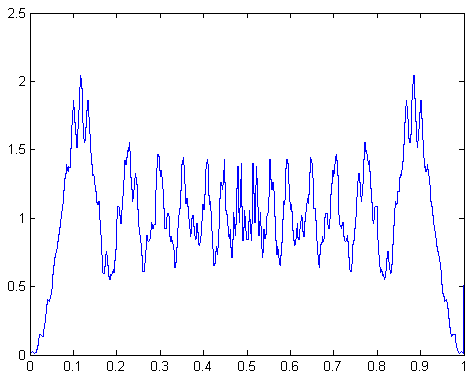} \\
\includegraphics[scale=0.39]{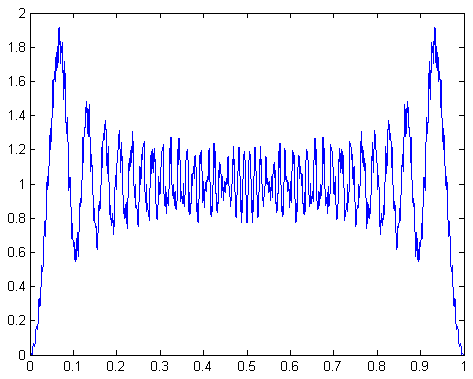} \includegraphics[scale=0.39]{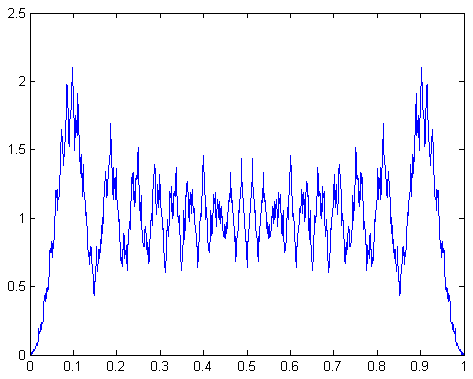} \includegraphics[scale=0.39]{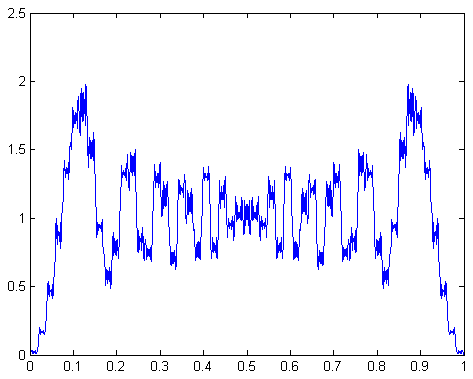}
\caption{Comparison of probability densities for a particle in a box as predicted by the quantum algorithm (top row) and from analytical solutions (bottom row), evolved from a ``flat'' distribution.  We use $N_t = 1000$, $n = 10$, and $L = 1$.  Time is in atomic units. \label{KE sim}}
\end{center}
\end{figure}

An exact wavefunction beginning in a uniform superposition can be expressed as a linear combination of eigenfunctions by Fourier expansion:
\[
\psi_\textrm{exact}(x, 0) = \frac{1}{\sqrt{L}} \hspace{2 mm} (0 < x < L) \implies \psi_\textrm{exact}(x, t) = \frac{2^{3/2}}{\pi}\sum_{k=1}^\infty \frac{1}{2k-1}\psi_{2k-1}(x)e^{-iE_{2k-1}t/\hbar}
\]
where $\psi_a = \sqrt{2/L}\sin(a\pi x/L)$, with corresponding energies $E_a = a^2\hbar^2\pi^2/2mL^2$.  We compare $|\psi_\textrm{exact}(x, t)|^2$ to the probability density of an initial input $|\psi(x, 0)\rangle = \frac{1}{2^{n/2}}\sum_{i=0}^{2^n-1}|i\rangle$ to the quantum algorithm as it evolves under the time evolution operators described in Section \ref{method}.  Figure \ref{KE sim} shows qualitatively the probability density of $\psi_\textrm{exact}(x, t)$ versus that of $|\psi(x, t)\rangle$ for several evolution times $T$.  In all simulations, 1000 terms are used in the Fourier series to compute $\psi_\textrm{exact}(x, t)$.

\begin{figure}[htb]
\begin{center}
\vspace{5 mm}
\includegraphics[scale=0.57]{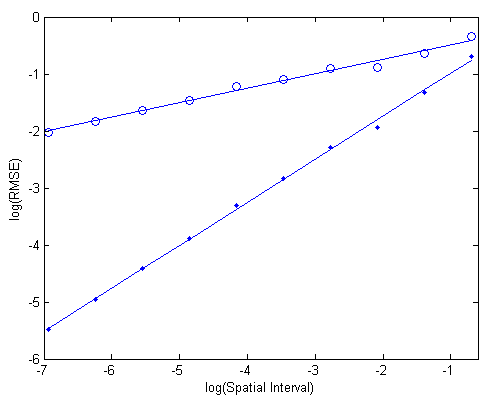} \includegraphics[scale=0.57]{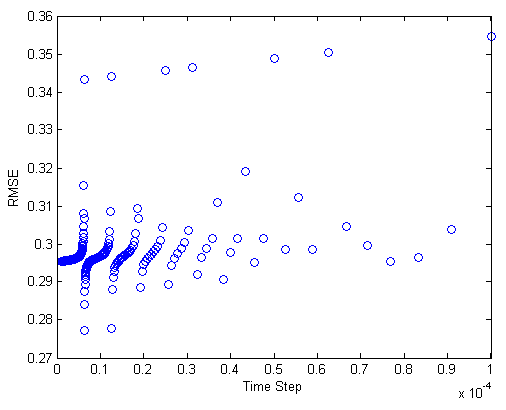}
\caption{Left: log-log plot of RMSE of $|\psi(x, t)|^2$ versus spatial interval $\delta$ at fixed $N_t = 1000$ for $n = 1, 2, \ldots, 10$.  Both $\log(\textrm{RMSE})$ and $\log(E_\textrm{YB})$ are shown as functions of $\log\delta$, with slopes of 0.2547 and 0.7547, respectively.  Right: RMSE versus time step $\epsilon$ at fixed $n = 6$ for $10\leq N_t\leq 1000$.  In both plots, $T = 10^{-3}$ and $L = 1$.  The error does not converge to 0 at spatial or temporal discretization 0 due to the fixed error in the other discretization. \label{convergence}}
\end{center}
\end{figure}

To estimate the convergence properties of the stated algorithm, we compute the root mean square error of the probability density of $|\psi(x, t)\rangle$ produced by the quantum algorithm versus both spatial and temporal discretization:
\[
\textrm{RMSE} = \frac{1}{2^{n/2}}\left(\sum_{i=0}^{2^n-1}\Big[\langle\psi(x, t)|\psi(x, t)\rangle - |\psi_\textrm{exact}(x, t)|^2\Big]^2\right)^{1/2}.
\]
The plots of Figure \ref{convergence} indicate that the error scales as a power of $\delta$, while preliminary results do not reveal the order of error in the temporal discretization.  We additionally compute the measure of error used by Yepez and Boghosian, defined by $E_\textrm{YB} = 2^{-n/2}(\textrm{RMSE})$ (\cite{Yepez}, eq. 30).  The slope of $\log(\textrm{RMSE})$ in the left plot is 0.2547 while that of $\log(E_\textrm{YB})$ is 0.7547, so the respective errors scale as $O(\delta^{0.2547})$ and $O(\delta^{0.7547})$.  This convergence in space is not nearly as efficient as with the QLGA representation of Schr\"odinger evolution, which usually converges with fourth-order error ($E_\textrm{YB}$) in $\delta$ \cite{Yepez}.  Thus future work might seek to increase the efficiency of our intuitive ``Cartesian'' operator representation.  From the right plot, the envelope of the maximum RMSE also appears to decrease slightly sublinearly with decreasing $\epsilon$, while interesting nonlinear patterns appear that suggest the presence of vertical asymptotes and more complicated behavior for most $\epsilon$.  Of course, the temporal error in the time evolution operator itself is second-order by equation \eqref{time evolution operator}, although this error can be reduced at the cost of increased gate complexity with a third-order Trotter decomposition.

Several subtle sources of numerical error arise in our simulations.  One downside to our construction of the kinetic energy operator is that its endpoint error, as mentioned in Section \ref{kinetic}, propagates with each application of the time evolution operator constructed from the approximate $\hat{P}$ in equation \eqref{P}.  However, this error affects at most a fixed small number of basis states in the particle state vectors at each time step.  As long as the spatial discretization sufficiently exceeds the temporal discretization, or $N_t\ll 2^n$, it should not noticeably hinder the simulation.  Figure \ref{KE sim} shows that even when $N_t\approx 2^n$, one obtains qualitatively excellent results: hence our construction of $\hat{P}$ as diagonal in the momentum basis is indeed robust.  Conversely, if $|h\rangle$ differs only slightly from an eigenvector of $\hat{H}$ (as a result of the approximation for $\hat{P}$), then $e^{ic\hat{H}}|h\rangle$ poorly approximates a phase shift of $|h\rangle$ for large $c$, so short absolute simulation times $T$ confer the best numerical stability.

Reasons that the RMSE does not converge exactly to 0 when either of the discretizations tend to 0 include the fixed error in the other variable, endpoint error, error due to a finite boundary potential, and (to a much lesser degree) error from a finite Fourier sum.



\subsection{Simple Molecules}

Though the electronic structure of molecules represents a physically interesting case, the present algorithm is limited in this domain because modeling electrons and nuclei as spinless precludes the simulation of atomic orbitals beyond the 1s. 

\begin{figure}[htb]
\begin{center}
\vspace{5 mm}
\includegraphics[scale=0.23]{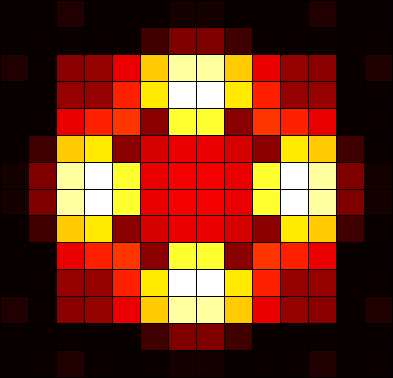} \includegraphics[scale=0.215]{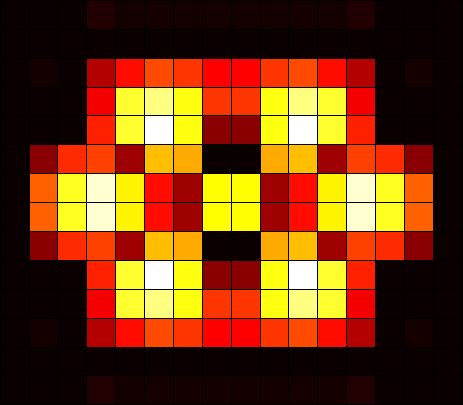}
\includegraphics[scale=0.229]{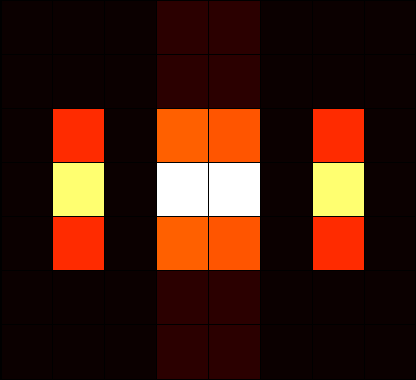} \includegraphics[scale=0.203]{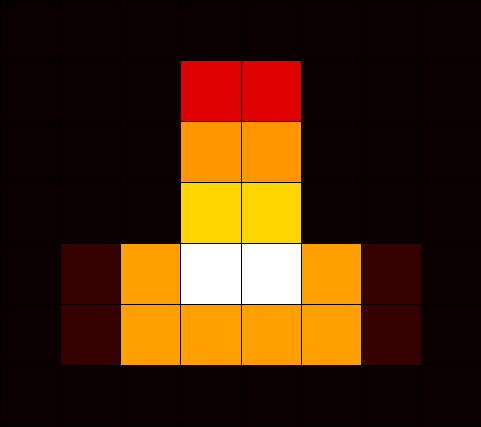}
\caption{2-D electron densities of H, H$_2^+$, H$_2$, and H$_3^+$, respectively, generated by the quantum simulation algorithm. \label{simSemiQM}}
\end{center}
\end{figure}

Nonetheless, the algorithm is simulated in MATLAB to determine what it would predict for the electronic charge density of several molecules, each with at most two spinless ``electrons'' and three nuclei.  To emulate the physical setup of a molecular system, the electron wavefunctions are initialized as uniform superpositions over subsets of position basis states via Hadamard transforms, while the much more massive nuclei are initialized to single position basis states.  For the range of precision that we consider, antisymmetrization makes little noticeable difference.  For simplicity, the numerical implementation uses fully quantum encoding and operators but the classical clamped-nucleus Born-Oppenheimer approximation.  We thus include only the terms $\hat{T}_e$, $\hat{U}_{ee}$, and $\hat{U}_{en}$ in the molecular Hamiltonian \eqref{molecular Hamiltonian}.  Also, we carry out the simulation in two dimensions both to reduce the space complexity and for easier visualization.  A normalized time interval of $T = 1$ is used.  The results of the algorithm for four very simple molecules are displayed in Figure \ref{simSemiQM}.  These plots show effective cross-sections of three-dimensional orbitals, where each square represents an electron position basis state.

\begin{figure}[htb]
\begin{center}
\vspace{5 mm}
\includegraphics[scale=0.21]{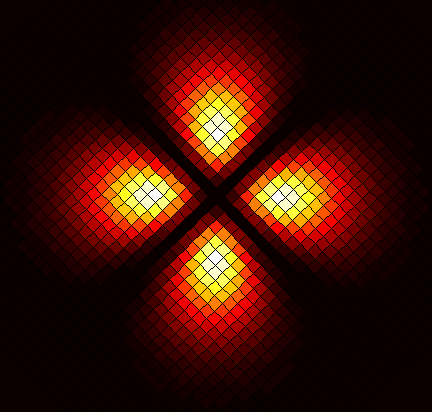} \includegraphics[scale=0.298]{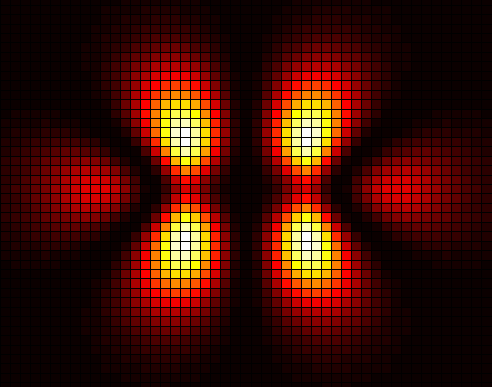}
\includegraphics[scale=0.267]{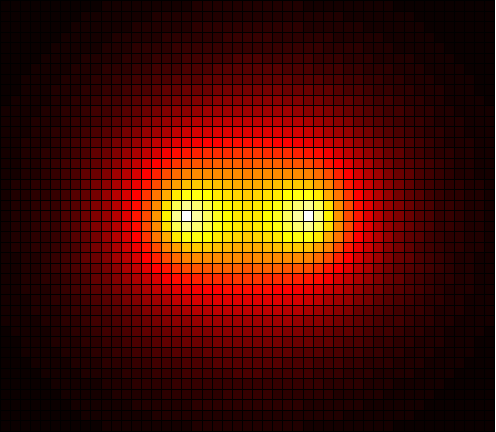} \includegraphics[scale=0.24]{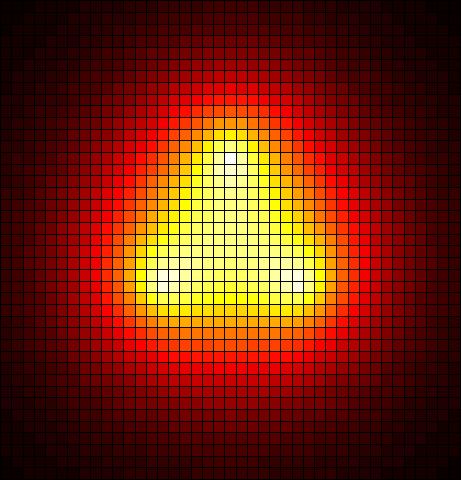}
\caption{Analytical and classical results for $|\psi|^2$ of H, H$_2^+$, H$_2$, and H$_3^+$ (only the latter two molecules are in the ground state).  The first image shows the exact electron density of an H atom corresponding to a 3d orbital in two dimensions, with wavefunction of the form $\psi(r_0, \theta) = \eta r_0^2 e^{-r_0/3}\sin 2\theta$.  H$_2^+$ is a linear combination of these exact 3d H orbitals.  The last two plots are outputs of classical variational methods. \label{classical simulations}}
\end{center}
\end{figure}

The very limited precision of the classical computer does not provide a detailed picture, but the plots all exhibit symmetry with respect to the nuclei.  To simulate H$_2$ and H$_3^+$ efficiently, the precision was decreased by a factor of four.  In these latter two plots, the greatest electron density appears to be in the center, where the atomic orbitals overlap.  The simulated hydrogen and H$_2^+$ orbitals, though lobed and artificially resembling those of a 3d subshell, most likely reflect numerical errors in the simulation due to discretization.  Compare the plots of Figure \ref{simSemiQM} to the classical results of Figure \ref{classical simulations}.

We have not addressed the question of simulating both ground and excited states of systems such as molecules.  Adjusting the input qubits to represent an initial configuration with a desired energy might allow simulation of states with arbitrary energies.  Alternatively, Zalka's proposed ``energy drain'' method could induce a simulated system to decay into its ground state by coupling it to an external reservoir \cite{Zalka}.

\section{Conclusion} \label{conclusion}

We have constructed the relevant quantum gates for simulating the non-relativistic Coulomb Hamiltonian for spinless many-body quantum systems, building on Wiesner and Zalka's model.  We have suggested implementations of these gates and proven the exactness of the kinetic energy operator in the continuum limit.  Lastly, we have simulated the algorithm in MATLAB and compared its outputs to analytical solutions, finding good agreement.  More numerical tests are needed and forthcoming.  In particular, it would be desirable to simulate two particles acting via the Coulomb force in a potential well in 1, 2, and 3 dimensions, and to compare those numerical tests involving interacting particles to classically obtained benchmarks.

Our preliminary investigation of this algorithm, under highly simplifying assumptions, has two main limitations.  First, disregarding fermionic spin statistics (the spin component of the wavefunction) means there is no way to enforce the Pauli exclusion principle, which does not currently permit the recovery of realistic atomic and molecular structure.  Second, if one chooses to sample the probability density of the many-body wavefunction at the end of the time evolution by making repeated runs of the algorithm and measuring (Section \ref{discretization}), the number of runs necessary to obtain sufficiently good ``resolution'' is unknown.  This latter problem is also encountered in the algorithm of Yepez and Boghosian \cite{Yepez}.

It should be possible to incorporate smaller terms into the Hamiltonian with no change in the form of the momentum operator.  Doing so lifts many of the simplifications we have made.  For example, the Dirac Hamiltonian corrects for both spin and relativistic effects \cite{Shankar}.  Two extra qubits are required to represent positive- and negative-energy states as well as spin up or down.  An electron wavefunction would thus reside in a Hilbert space $\mathcal{H}_r \otimes \mathcal{H}_s^{\otimes 2}$ where $\mathcal{H}_r$ is $2^{3n}$-dimensional and $\mathcal{H}_s$ is two-dimensional.  This results in four-component spinor wavefunctions where each component is itself a position-dependent wavefunction.  The Dirac Hamiltonian for an electron in an electromagnetic field is
\[
\hat{H}_D = \gamma^0(mc^2 + \gamma^\mu \pi_\mu c) + \hat{U}
\]
where $\gamma^\mu$ are the gamma matrices and $\vec{\pi} \equiv \vec{P} + e\vec{A}/c$ where $\vec{P}$ and $\vec{A}$ are operator-valued vectors of momentum operators and vector potential operators, respectively, in the $x$, $y$, and $z$ directions.  The momentum operators in each of the three spatial dimensions act only on the qubits encoding the electron's position in that dimension.  Namely, $\vec{P} = (\hat{P}_x, \hat{P}_y, \hat{P}_z)$ with $\hat{P}_x = \hat{P} \otimes I_{2^{2n}}$, $\hat{P}_y = I_{2^n} \otimes \hat{P} \otimes I_{2^n}$, $\hat{P}_z = I_{2^{2n}} \otimes \hat{P}$, and $\hat{P}$ as defined in equation \eqref{P}.  Because the gamma matrices act on the spinor components, we can rewrite these operators in terms of matrices of the correct dimension as
\[
\hat{H}_\textrm{relativistic + spin} = \gamma^0 mc^2 \otimes I_{2^{3n}} + (\gamma^0 \gamma^\mu \otimes I_{2^{3n}})(I_4 \otimes \pi_\mu c) + I_4 \otimes \hat{U}.
\]
Both the position and the velocity of every particle in the system make a nonlocal contribution to the vector potential at every point, so implementing $\vec{A}$ might therefore require additional qubits for keeping track of such velocities.  Finally, these corrections we have described account only for electron spin and not nuclear spin.

\section*{Acknowledgements}

The author thanks Dr. Marek Perkowski of the Portland State University ECE Department and Dr. Pui-Tak Leung of the PSU Department of Physics for discussions, as well as Jacob Biamonte of the Oxford University Computing Laboratory for helpful references.

\end{document}

%% file: paperformats.tex
\newcommand{\arxivformat}
{
\addtolength{\hoffset}{-0.5 in} \addtolength{\textwidth}{1 in}
\addtolength{\voffset}{-0.75 in} \addtolength{\textheight}{1 in}
}

